\documentclass[aps,prb,twocolumn,showpacs]{revtex4}

\usepackage{epsfig}

\usepackage{afterpage}

\usepackage{epsfig,amsmath}

\usepackage{color}

\usepackage{amsmath}

\usepackage{amssymb}

\usepackage[latin1]{input enc}

\usepackage{graphics}

\usepackage{graphicx}

\usepackage{subfigure}

\usepackage{psfrag}

\begin{document}

\newcommand{\nn}{\nonumber}

\newcommand{\bb}{\begin{eqnarray}}

\newcommand{\ee}{\end{eqnarray}}

\newcommand{\Vol}{{\cal A}}

\newcommand{\sign}{{\rm sign}}

\newcommand{\LL}{\textsf{L} }

\newcommand{\Ar}{{\cal A}_0}

\newcommand{\ar}{\sigma}

\newcommand{\ff}{\frac{1}{2}}

\newcommand{\Tc}{\tilde T}

\newcommand{\Bc}{\tilde B}

\newcommand{\Ec}{\tilde E}

\newcommand{\amp}{A_1}

\title{\textbf{Damping of field-induced chemical potential oscillations in ideal two-band compensated metals}}

\author{Jean-Yves Fortin$^1$}\email{fortin@lpt1.u-strasbg.fr}

\author{Alain Audouard$^2$}\email{audouard@lncmp.org}

\affiliation{ $^1$Laboratoire de Physique Th\'eorique,
Universit\'e Louis Pasteur (UMR CNRS-ULP 7085), 3 rue
de l'Universit\'e, F-67084 Strasbourg cedex, France\\
$^2$Laboratoire National des Champs Magn\'etiques Puls\'es (UMR
CNRS-UPS-INSA 5147) 143 avenue de Rangueil, F-31400 Toulouse,
France}

\date{\today}

\begin{abstract}

The field and temperature dependence of the de Haas-van Alphen
oscillations spectrum is studied for an ideal two-dimensional
compensated metal. It is shown that the chemical potential
oscillations, involved in the frequency combinations observed in
the case of uncompensated orbits, are strongly damped and can even
be suppressed when the effective masses of the
electron- and hole-type orbits are the same. When magnetic breakdown between
bands occurs, this damping is even more pronounced
and the Lifshits-Kosevich formalism accounts for the data in a wide field range.

\end{abstract}


\maketitle

\section{Introduction}

In large enough magnetic field, the Fermi surface (FS) of
multiband quasi-two-dimensional metals, is liable to give rise to
networks of orbits coupled by magnetic breakdown (MB). The most
studied type of network is the linear chain of coupled orbits
introduced by Pippard \cite{Pi62} and illustrated by several
quasi-two-dimensional (q-2D) organic conductors such as
$\kappa$-(BEDT-TTF)$_2$Cu(NCS)$_2$. As discussed, in Ref.
\cite{Ka04rev}, magnetic oscillations spectra of such
networks contain many frequencies that are linear combinations of
two basic frequencies. In addition to those linked to MB-induced
orbits, other frequencies are observed that are not accounted for
by the semi-classical theory of Falicov and Stachowiak
\cite{Fa66}. They can be attributed to quantum interference (as
far as magnetoresistance oscillations are concerned), MB-induced
modulation of the density of states \cite{Sa97, Fo98, Gv02} and
oscillation of the chemical potential \cite{Al96,Ch02,Ki02,Fo05},
even though the actual respective contribution of these three phenomena to the oscillatory behavior remains to be established.\\

\begin{figure}
\centering
\resizebox{1\columnwidth}{!}{\includegraphics*{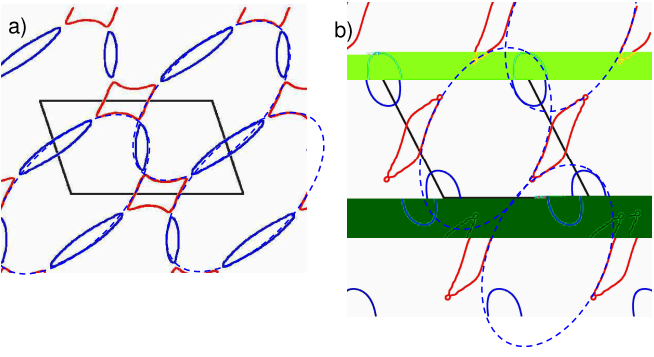}}
\caption{\label{fig_Networks} (color online) Calculated Fermi
surface (FS) of q-2D organic metals (a)
(BEDO)$_4$Ni(CN)$_4\cdot$4CH$_3$CN \cite{Du05} and (b)
(BEDT-TTF)$_4$NH$_4$[Fe(C$_2$O$_4$)$_3$] \cite{Pr03} leading to
networks of compensated electron- and hole-orbits (solid red and
blue lines, respectively). Ellipses in dotted blue lines
correspond to the hole orbits, with an area equal to that of the
FBZ, from which the FS is built (see text).}
\end{figure}

Another type of network is provided by q-2D metals of which the 
FS is composed of compensated electron-
and hole-types tubes. This is the case of the
family of organic metals
(BEDT-TTF)$_8$Hg$_4$Cl$_{12}$(C$_6$H$_5$X)$_2$ (X = Cl, Br) whose
FS, which originates from two pairs of crossing q-1D sheets, is
composed of one electron and one hole tube with the same area
\cite{Ve94}. As it is the case of the above-mentioned linear
chains of coupled orbits, magnetoresistance oscillations spectra
in this type of network reveal frequencies that are linear
combinations of three basic frequencies, linked to the compensated
orbits and to the two FS pieces located in-between \cite{Pr02,
Vi03, Au05}. However, in striking contrast to the data relevant to
linear chains of orbits, de Haas-van Alphen (dHvA) oscillations
spectra recorded in the case of the compound with X = Br only
exhibit oscillations, the field and temperature dependence of
which can be consistently interpreted on the basis of the
semiclassical model of Falicov and Stachowiak \cite{Au05,Fa66}.
Analogous networks are observed in organic metals, with two
carriers per unit cell. In this case, the FS originates from the
overlapping in two directions of hole tubes with an area equal to
that of the First Brillouin zone (FBZ) and from the resulting gap
openings \cite{Ro04}. As reported in the case of
(BEDO)$_4$Ni(CN)$_4\cdot$4CH$_3$CN \cite{Du05}, such a FS yields a
network consisting in two hole- and one electron-type tubes (see
Fig. \ref{fig_Networks}(a)). Closely related network is obtained
in the case where the large hole orbit come close to the FBZ
boundary, as it is the case of
(BEDT-TTF)$_4$NH$_4$[Fe(C$_2$O$_4$)$_3$] \cite{Pr03}. In this
latter case, a large MB gap is observed at this point and the
resulting network only consists in one electron- and one hole-type
orbit, as displayed in Fig. \ref{fig_Networks}(b).
Linear chain of successive electron-hole tubes might also be observed
in Bechgaard salt (TMTSF)$_2$NO$_3$ \cite{PRL50p1005,PRB020404}.

The aim of this paper is to explore the field and temperature
dependence of the dHvA oscillations spectra of an ideal 2D metal
whose FS is composed of one electron- and one hole-type
compensated orbit. It is demonstrated that the field-induced
oscillations of the chemical potential are strongly damped for such
a FS and can even be suppressed in the case where electron- and
hole-type orbits have the same effective masses. The chemical
potential oscillations can be even more damped when the two orbits
are coupled by MB. In this case, the Lifshits-Kosevich (LK)
formalism accounts for the data up to high magnetic field and low
temperature.

\section{Model}

We first consider a 2D metal whose electronic structure consists of two
parabolic bands of hole- and electron-type, respectively. The
bottom of the electron  band is set at zero energy while the top
of the hole band is at $\Delta > 0$. The total number of electrons
in the system is such that the hole band is completely filled.
Since the lower part of the electron band is lower in energy than
the top of the hole band, some quasiparticles move to the electron
band in order to lower the total energy. The effective masses
linked to the two bands, $m_e^{*}\equiv 1/C_e$ and $m_h^{*}\equiv
1/C_h>0$, can be different. It is useful to define the physical
units of the problem. The reduced field $b=eB\Ar/h$ is in units of
the characteristic field $\Bc=h/e\Ar$, the effective masses are in
units of the electron mass $m_e$, the energies are in units of
$\Ec=2\pi\hbar^2/m_e\Ar$ and the temperature $t$ in units of
$\Tc=\Ec/k_B$. Given a unit cell area $\Ar=$197.6 $\AA^2$, which
stands for (BEDT-TTF)$_8$Hg$_4$Cl$_{12}$(C$_6$H$_5$Cl)$_2$
\cite{Ly91}, we obtain $\Bc = 2093$ T and $\Tc = 2812$ K.
Therefore, realistic experimental conditions yield small values of
$b$ and $t$ compared to $\Bc$ and $\Tc$, respectively. The
semi-classical quantization of the energy levels in the presence
of a magnetic field leads to the Landau equations:

\bb\label{LL0}
E_e(n)=C_eb(n+\ff),\;\;\;E_h(n)=\Delta-C_hb(n+\ff)
\ee

each Landau level (LL) having a degeneracy $b$ per sample area.
The zero field Fermi energy is simply given by
$E_F=m_h^{*}/(m_e^{*}+m_h^{*})\Delta$. At finite temperature, the
total free energy is given by the difference between the
contribution of the electron and of the hole band, with the
condition that the total number of quasiparticles $N_{eh}$ is fixed.
In addition, the compensation condition imposes that the number of
quasiparticles in the electron ($N_e$) and in the hole ($N_h$)
band are the same. From the thermodynamical relations, we can
define a free energy for the system based on the difference
between the free energy for the electrons in the electron band and
the free energy for the holes in the hole band $\Delta
F=\Omega_e-\Omega_h+(N_e-N_h)\mu$, where $\Omega_{e(h)}$ is the
Grand Potential for the electrons (holes) and $\mu$ is the
chemical potential [$\mu(t=0,b)=E_F(b)$]:

\bb \nn \Omega_e(t,b)&=& -tb\sum_{n\ge 0}\log\left
(1+\exp[\beta(\mu-E_e(n))]\right )\\
\label{omega} \Omega_h(t,b) &=& tb\sum_{n\ge 0}\log\left
(1+\exp[\beta(E_h(n)-\mu)]\right ) \ee

Since $N_e=N_h$, we conclude that $F=\Omega_e-\Omega_h$ in
compensated metals. $\mu$ is evaluated from the self-consistent
equation $\partial \Delta F/\partial\mu=0$. At zero temperature,
the above expressions reduce simply to the ground state (GS)
energy $\Delta E_0$. The Fermi energy $E_F(b)$ is given by the
condition $n_eb=n_hb$, where $n_e$ and $n_h$ are the (integer) numbers of LL
filled, and

\bb \nn \Delta
E_0&=&-b\sum_{n=0}^{n_e-1}(E_F(b)-E_e(n))-b\sum_{n=0}^{n_h-1}(E_h(n)-E_F(b))
\\ \label{GS}
&=&E_e-E_h+ b(n_h-n_e)E_F(b)=E_e-E_h \ee

The special cases where the Fermi energy goes trough one Landau
level, or where this Landau level is partially filled, correspond
to singular points in the energy spectrum as a function of the
inverse field that do not modify the thermodynamical quantities.
The exact expression for $\Delta E_0$ is simply

\bb\label{E0}
\Delta E_0 = \frac{1}{2}(C_e+C_h)b^2n_e^2-b\Delta n_e \ee


with $n_e=[E_F/C_eb+1/2]_i$, the notation $[.]_i$ standing for the
integer part of the argument. The GS energy oscillates around the
limit of zero field, where $\Delta E_0(q=1)=-\Delta^2/2(C_e+C_h)$.
For example, taking $\Delta=1$, $C_e=1$ and $C_h=2/5$, we obtain
$\Delta E_0\simeq-0.357$. We deduce the oscillating part of the
magnetization $M_{osc}=-\partial \Delta E_0/\partial b$  from the
latter expression, using succesively the Fourier transforms of the periodic
functions $[x]_i-x$ and $([x]_i-x)^2$:

\bb M_{osc}=-(C_e+C_h)F_0\sum_{k\ge 1}\frac{(-1)^k}{\pi k} \sin
(2\pi k F_0/b) \ee

where
$F_0=\Delta/(C_e+C_h)=m_e^{*}m_h^{*}\Delta/(m_e^{*}+m_h^{*})$ is
the fundamental frequency corresponding to the FS area of the
electron and hole band. At zero temperature and for fixed number
of electrons, the magnetization oscillates with characteristic
frequency $F_0$, and the amplitudes $A_k$ of the $k^{th}$
harmonics are given by the LK formula with $1/k$ dependence
\cite{shoenberg}:

\bb
A_k=(-1)^{k+1}\frac{F_0}{k\pi}(C_e+C_h)
\ee

The sum $C_e+C_h$ means that the 2 orbits circulating around the hole and
electron bands contribute individually to the magnetization.

\section{Self-consistent equation for the chemical potential}

The oscillatory parts of the Grand Potentials Eqs. (\ref{omega}) can be extracted
using Poisson's formula for any function $F(x)$ (see Ref. \cite{shoenberg,LK} for details):

\bb \nn
& &\frac{1}{2}F(0)+\sum_{n\ge 1}F(n)=\int_0^{\infty}
F(x)dx
\\ \label{poisson}
&+&2\Re \sum_{n\ge 1}\int_0^{\infty}
F(x)\exp (2i\pi nx)dx
\ee

where $F(n)$ is equal to $\log(1+\exp[\beta(\mu-E_e(n))])$ or
$\log(1+\exp[\beta(E_h(n)-\mu)])$ for electrons and holes,
respectively. The last series in Eq. (\ref{poisson}) gives the
oscillatory part of the Grand Potential in terms of Fourier
components. The index $n$ is expressed as a function of the energy $n=n(E)$
by relations (\ref{LL0}), then expanded around the chemical
potential $\mu$ at low temperature where the energy derivatives of
the distribution functions $1/(1+\exp[\beta(E_e(n)-\mu)])$ or
$1/(1+\exp[\beta(\mu-E_h(n))])$ are strongly peaked. After some algebra, we
obtain for each band

\bb \label{Omega_e} \Omega_e&\simeq&-\frac{1}{2C_e}\mu^2
\\ \nn
&+&\frac{b^2C_e}{2}
\left [
\frac{1}{12}+\sum_{n=1}^{\infty}\frac{(-1)^n}{\pi^2n^2}R(nm_{e}^*)\cos(2\pi
n\frac{\mu}{C_eb}) \right ]
\\ \label{Omega_h}
\Omega_h&\simeq&\frac{1}{2C_h}(\Delta-\mu)^2
\\ \nn
&-&\frac{b^2C_h}{2} \left [
\frac{1}{12}+\sum_{n=1}^{\infty}\frac{(-1)^n}{\pi^2n^2}R(nm_{h}^*)\cos(2\pi
n\frac{\Delta-\mu}{C_hb}) \right ] \ee

where

\bb
R(nm_{e(h)}^*)=\frac{2\pi^2nm_{e(h)}^*t/b}{\sinh(2\pi^2nm_{e(h)}^*t/b)}
\ee

 is the temperature reduction factor for effective masses $nm_{e(h)}^*$.
A Dingle term
$R_D(nm_{e(h)}^*,t_{e(h)}^*)=\exp(-2\pi^2nm_{e(h)}^*t_{e(h)}^*/b)$
can also be added in the case where the relaxation times $t_e^*$
and $t_h^*$ for electron- and hole-band have to be taken into
account. In the following we will assume that these two relaxation
times are negligible for convenient purpose. It is always
possible to add those terms in final expressions. The chemical
potential satisfies therefore the self-consistent relation:

\bb \nn \mu=E_F&+&\frac{b}{m_e^*+m_h^*}
\sum_{n=1}^{\infty}\frac{(-1)^n}{\pi n} [ R(nm_{h}^*)\sin(2\pi
n\frac{\Delta-\mu}{C_hb})
\\ \label{muexact}
&-&R(nm_{e}^*)\sin(2\pi n\frac{\mu}{C_eb}) ]
\ee

At a first order approximation, which will be discussed in the
following sections, we can replace $\mu$ in the sine functions of the
previous expression by $E_F$ when the oscillations of the chemical
potential are small compared to $E_F$, in the small field and high
temperature regime ($t/b$ large). This gives:

\bb \label{mu} \mu&\simeq& E_F+\frac{b}{m_e^*+m_h^*}\times
\\ \nn
& &\sum_{n=1}^{\infty}\frac{(-1)^n}{\pi n} [R(nm_{h}^*)-R(nm_{e}^*)]\sin(2\pi
n\frac{F_0}{b})
\ee


\begin{figure}
\centering
\resizebox{0.85\columnwidth}{!}{\includegraphics*{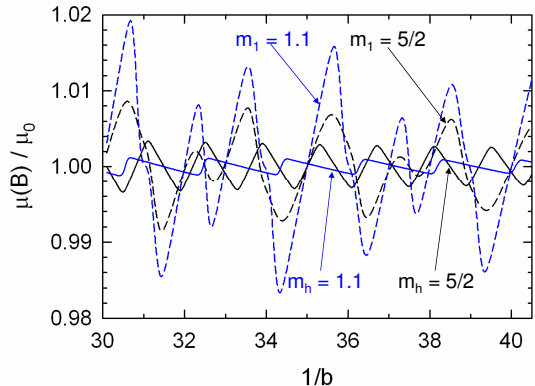}}
\caption{\label{fig_mu} (color online) Field dependence of the
chemical potential normalized to its value in zero-field at t =
0.001. Solid and dashed lines correspond to one electron and one
hole compensated orbits (with m$^*_e$ = 1 and m$^*_h$ = 1.1 or
5/2) and two electron orbits (with m$^*_0$ = 1 and m$^*_1$ = 1.1
or 5/2), respectively (see text). }
\end{figure}

It can be remarked first that, in the case where the effective
masses and relaxation times linked to the electron- and hole-type
orbits are the same, there is an exact solution for Eq.
(\ref{muexact}) with $\mu=E_F$. In this case the oscillatory part
of Eq. (\ref{muexact}) or Eq. (\ref{mu}) vanishes and the chemical
potential remains constant in magnetic field and temperature. This
is due to the fact that the energy levels are symmetric around
$E_F$. More generally, the amplitude of the chemical potential
oscillations can be compared to the case of two electron bands
\cite{Fo05}. In Fig. \ref{fig_mu}, the field-dependent chemical
potential Eq. (\ref{muexact}) is calculated for compensated orbits with m$^*_e$ = 1
and m$^*_h$ = 1.1 or 5/2 and compared to the case of two
electronic orbits with effective masses m$^*_0$ = 1 and m$^*_1$ =
1.1 or 5/2. It can be observed that the chemical potential
oscillations are strongly damped for compensated orbits, even in
the case where m$_h^*$ and m$_e^*$ have strongly different values
\cite{fn1}.

\section{De Haas-van Alphen oscillations}

\begin{figure}


\centering
\resizebox{0.8\columnwidth}{!}{\includegraphics*{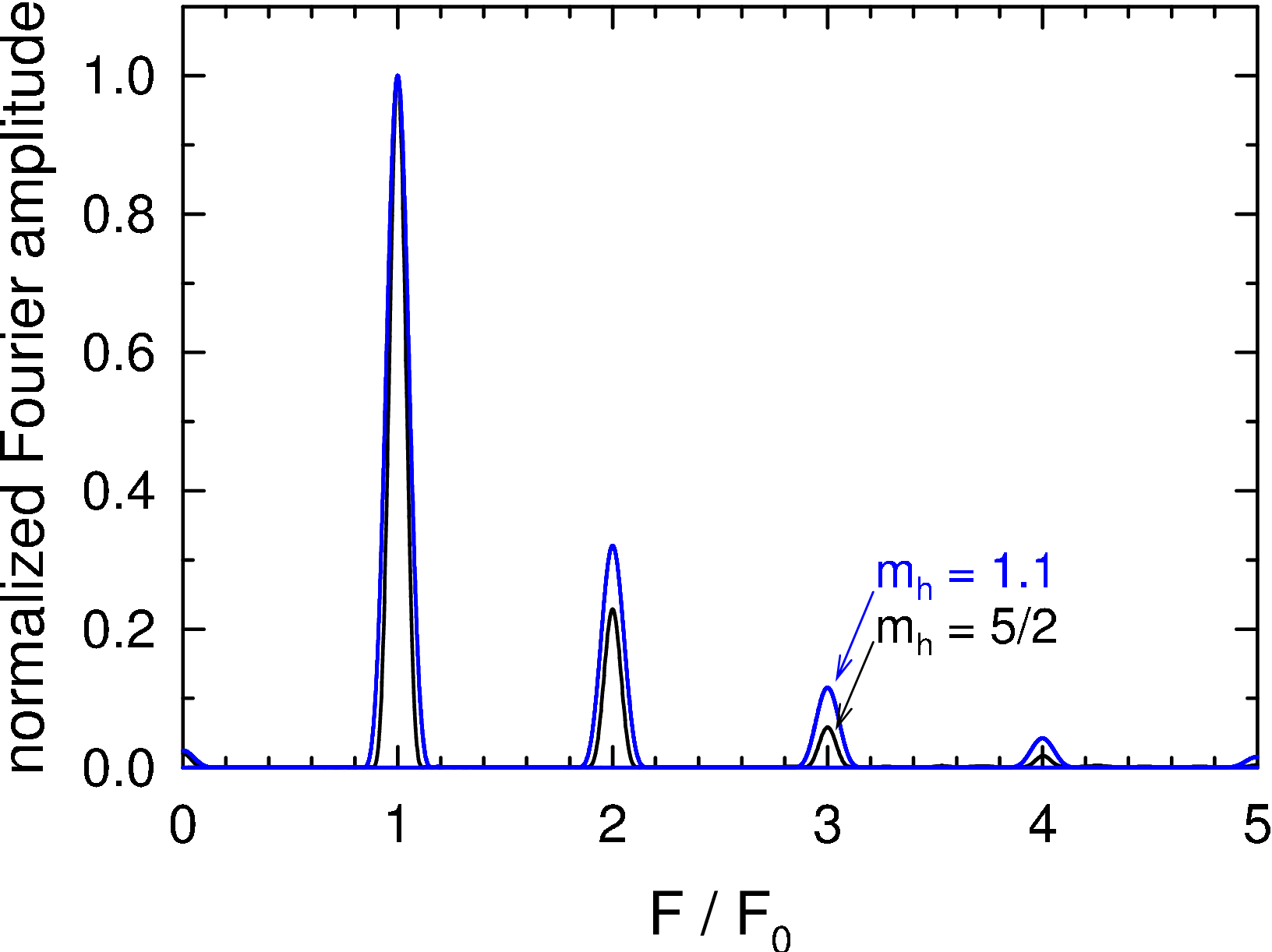}}
\caption{\label{fig_TF} (color online) Fourier spectrum of the
magnetization at t = 0.001 for m$^*_e$ = 1 and m$^*_h$ = 1.1 or
5/2. The field range is from b = 0.015 to b = 0.033. F$_0$ is the
fundamental frequency (see text). }
\end{figure}

The oscillatory magnetization can be obtained putting the
solution of the chemical potential given by Eq. (\ref{muexact}) back
into the expression for the free energy $F$ using Eqs.
(\ref{LL0}) and (\ref{omega}). As discussed above, the chemical potential is
field-independent for m$_e$ = m$_h$ and the LK
formula \cite{shoenberg} holds in this case. However, for m$^*_e
\neq$ m$^*_h$, the oscillatory magnetization differs from the LK
theory. Examples of Fourier analysis of the magnetization are
given in Fig. \ref{fig_TF}. It should be noted that, contrary to the case of electron-hole systems away from exact compensation \cite{Gv07},
frequency combinations are not observed directly in the oscillatory spectra.
However, as discussed later on, deviations from the LK behavior are observed for the second harmonics.
\\

\begin{figure}
\centering \resizebox{1.1\columnwidth}{!}
{\includegraphics*{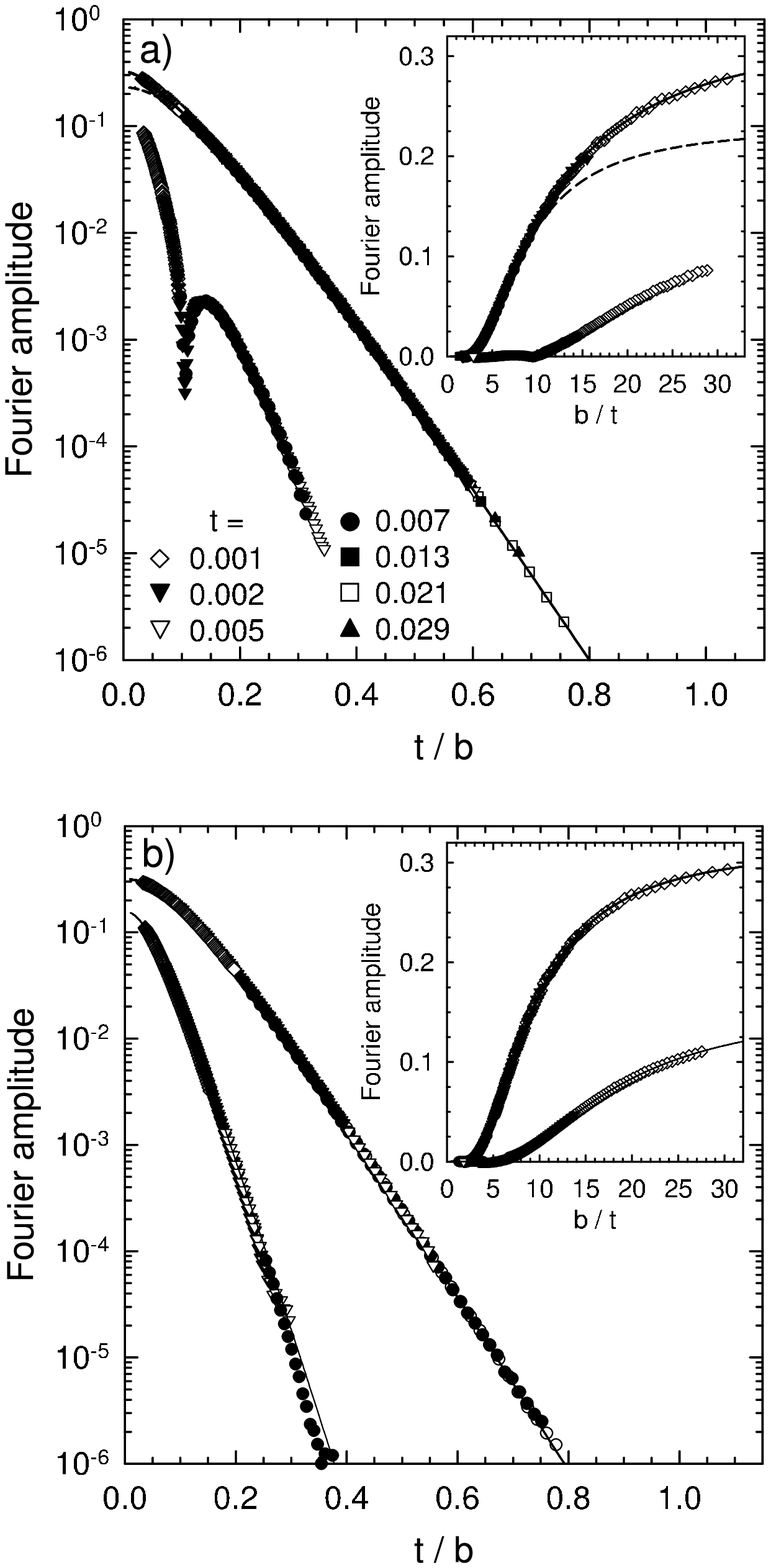}} \caption{\label{fig_dingle} t/b
dependence of the Fourier components of the magnetization with
frequencies F$_0$ and 2F$_0$. The effective mass of the
electronic-type orbit is m$_e^*$ = 1. The effective mass of the
hole-type orbit is m$_h^*$ = 5/2 and 1.1 in (a) and (b),
respectively. Solid lines are obtained from the LK formula. The
dashed line in (a) corresponds to the contribution of the
electronic orbit, only.}
\end{figure}

It is useful for experimentalists, to check to what extent it is
possible to determine a temperature and field range in which the
LK formalism provides a satisfactory approximation of the
oscillatory behavior in the case where m$^*_e \neq$ m$^*_h$. The
b/t dependence of the Fourier components of the magnetization with
frequencies F$_0$ and 2F$_0$ are given in Fig. \ref{fig_dingle}.
The LK formula accounts for the field and temperature dependence
of the first harmonics. Furthermore, it can be noticed that in the
case where m$_h^*$ is strongly different from m$_e^*$, the
contribution of the orbit with the smallest effective mass
dominates in a large field range (see the dashed line in Fig.
\ref{fig_dingle}a). It has also been checked that, in the opposite
case where m$^*_e$ is close to m$^*_h$, the data can be accounted
for by the contribution of only one orbit with a mean effective
mass, namely, m$^*_{mean}$ $\simeq$ (m$^*_e$ + m$^*_h$)/2. In
contrast, the LK formula cannot account for the second harmonics
in the case where m$^*_e$ and m$^*_h$ significantly differ. The
observed behavior of the second harmonics in this latter case is
reminiscent to that observed for two electronic orbits
\cite{Fo05}.

\section{Magnetic Breakdown}

In this section, we consider the case where the magnetic field is
large enough for the quasiparticles can tunnel through MB between
the two bands.

\begin{figure}
\centering \resizebox{\columnwidth}{!}
{\includegraphics{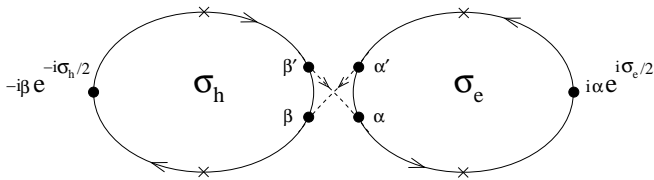}}
\caption{\label{FSMB} Fermi surface of a two-band model with
magnetic breakdown in the FBZ. The quasiparticles orbit
clockwise for the Hole surface and counterclockwise for the Electron band. They can tunnel
between the points (which represent also the amplitudes of the wave-function)
$\alpha'$ and $\beta$ (or $\beta'$ and $\alpha$)
with a probability amplitude $ip$, or be relected between $\alpha'$ and $\alpha$
(or $\beta'$ and $\beta$) with a probability amplitude $q$. At the turning points (cross
symbols), the wavefunction acquires a factor $i$ (or $-i$) depending wether the quasiparticle orbits an electron or hole surface.}
\end{figure}

The topology of the new FS is depicted in Fig. \ref{FSMB}: the
quasiparticles can tunnel between the junctions $\alpha'$ and
$\beta$ ($\beta'$ and $\alpha$) with probability amplitude $ip$ or
be reflected between the junctions $\alpha'$ and $\alpha$
($\beta'$ and $\beta$ respectively) with probability amplitude
$q$. This surface is a simplification of Fig. \ref{fig_Networks}b
in the sense that it does not include the 1D network feature.
$p$ and $q$ are related by $p^2 + q^2 = 1$
\cite{Fa66,shoenberg}, and the field dependence of $p$ is given by
the Chambers formula $p=\exp(-b_0/2b)$ \cite{chambers,kadi}, where
$b_0$ is a characteristic MB field. $b_0$ comes in the field range
covered by experiments in the case where the two bands are located
closely enough in the FBZ. Depending on the band type
(electron or hole), the quasiparticle wave-function acquires a
phase either $\ar_e=S_e/b$ or $\ar_h=S_h/b$ when orbiting around
each part of the Fermi surface (between $\alpha$ and $\alpha'$ counterclockwise for
the electron band or $\beta$ and $\beta'$ clockwise for the hole band). The
semiclassical actions $S_e=2\pi m_e^{*} E$ and $S_h=2\pi m_h^{*}
(\Delta-E)$ represent the areas delimited by each Fermi surface in
the FBZ. Assuming that the wavefunction of the
quasiparticle is $\alpha$ at the point $\alpha$, we obtain that after one 
orbit around the electron band, $\alpha'=-\alpha\exp(i\sigma_e)$ (the quasiparticle
goes through 2 turning points where it acquires a factor $i$ or
$-i$ each time, depending on the electron or hole type
\cite{kadi}). Similarly, it follows $\beta'=-\beta\exp(-i\sigma_h)$ for the hole band. The four amplitudes $\alpha$, $\alpha'$, $\beta$ and $\beta'$ are related
by the transfer matrix \cite{kadi,Fo98}

\bb
\left (
\begin{array}{c}
\alpha \\ \beta
\end{array}
\right )
=
\left (
\begin{array}{cc}
q & ip
\\
ip & q
\end{array}
\right )
\left (
\begin{array}{c}
\alpha' \\ \beta'
\end{array}
\right )
\ee

This set of equations have non-zero solution when the energy $E$
satisfies the implicit equation:

\bb \nn (1 + q\exp i\ar_e)(1 + q\exp(-i\ar_h))\\
\label{LLq}+ p^2\exp i(\ar_e-\ar_h)=0
\ee

\begin{figure}
\centering \resizebox{\columnwidth}{!}
{\includegraphics*{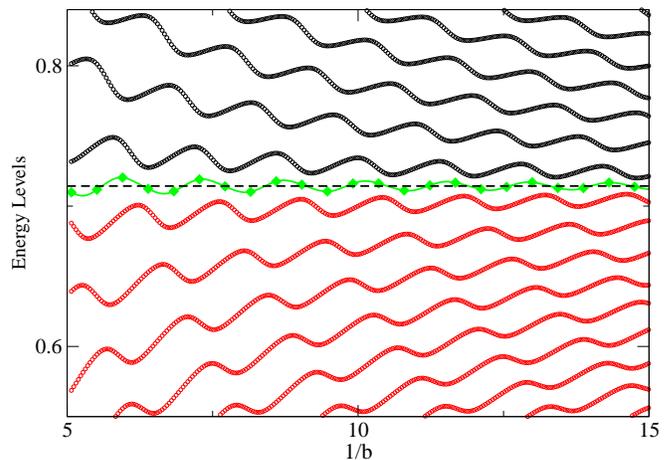}}
\caption{\label{Levelq06me1mh52} (color online) Energy levels for a
reflection amplitude $q = 0.6$ (or, equivalently, tunnelling
amplitude $p = \sqrt{1 - q^2}$), $\Delta = 1$, m$_e^*$ = 1 and
m$_h^*$ = 5/2. Levels that belong to a quasi-electron band (going
upwards in the limit of large field, black lines) above the Fermi
energy $E_F(b)$ (diamond symbols, green line) can be distinguished from levels that belong to a
quasi-hole band (respectively going downwards, red lines). The Fermi energy
at zero field is shown as a dashed line.}
\end{figure}

\begin{figure}
\centering \resizebox{\columnwidth}{!}
{\includegraphics*{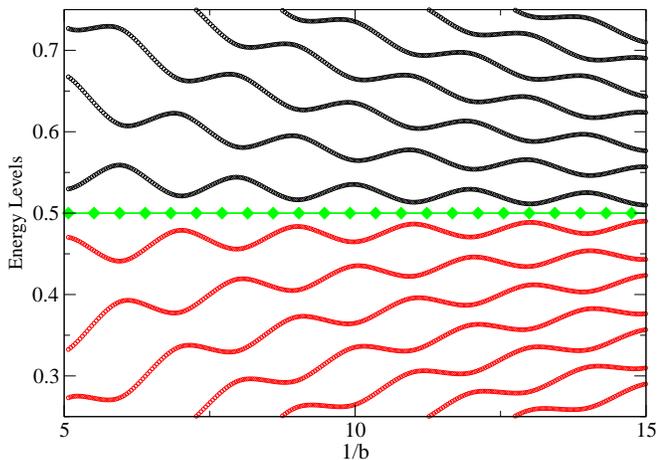}}
\caption{\label{Levelq06me1mh1} (color online) Energy levels for a
reflection amplitude $q = 0.6$, $\Delta = 1$, m$_e^*$ = m$_h^*$ = 1.
Levels that belong to a quasi-electron band above the Fermi
energy located at $E_F(b)$=0.5 (diamond symbols, green line) can be distinguished from
levels that belong to a quasi-hole band (red lines).
The 2 bands are fully symmetric with respect with the chemical potential energy.}
\end{figure}

In the case where the bands are disconnected ($p = 0$),  Eq.
(\ref{LLq}) can be factorized and reduces simply to $1 +
\exp(i\ar_e) = 0$ or $1 + \exp(-i\ar_h) = 0$, which corresponds to
the two independent sets of discrete energy levels given by Eq.
(\ref{LL0}). In general Eq. (\ref{LLq}) can be solved numerically
for $p \neq 0$: the field dependence of the LLs energy is plotted
in Figs. \ref{Levelq06me1mh52} and \ref{Levelq06me1mh1} for
$q$=0.6 and for 2 different sets of effective masses. In the case
where $p\ne 0$, gaps open at the former LL intersections, and the
structure of the two bands is modified. Alike in the case where $p
= 0$, a quasi-hole and a quasi-electron band can be defined.
Indeed, increasing the magnetic field, levels that go upwards
above the Fermi energy $E_F(b)$ (see diamond symbol lines) and
downwards below $E_F(b)$ represent quasi-electron and quasi-hole
bands, respectively. At zero temperature, the compensation
condition $N_e = N_h$ is replaced by the condition that the lower
quasi-hole band is filled completely, the upper electron-band
being empty. The two bands are therefore always separated by the
chemical potential energy, as shown in Figs. \ref{Levelq06me1mh52} and
\ref{Levelq06me1mh1}. As the magnetic field decreases, the
field-dependent Landau levels near the Fermi energy become flatter
since the levels of each band do not intersect the Fermi level. In
the case where $q=0$ ($p=1$), the quasiparticles tunnel directly
through the 2 bands at each revolution, and the exact solution of
Eq. (\ref{LLq}) is given by the set of LLs:

\bb\nn
E_{eh}(q=0,n)&=&E_F-b\frac{C_eC_h}{C_e+C_h}(n+\ff)
\\ \label{LLq0}
&=&\frac{C_e}{C_e+C_h}E_h(n)
\ee

with $n\ge 0$. Within the semi-classical approach, the
cross-section area of the FS corresponding to this inverted
parabolic quasi-hole band, which is completely filled, is zero.
Since it is proportional to the frequency of the oscillations,
there are no oscillations at $q = 0$. This simple feature is in
line with the prediction of the semiclassical model for two
compensated orbits, keeping in mind that the area of these orbits
has opposite signs \cite{Fa66}.

We assume that $C_h=C_ep_0/q_0$ where $p_0$ and $q_0$ are coprime integers \cite{Fo98,Fo05}. Indeed, it is useful, also for computing time reasons, to approximate any real value of $C_h/C_e$ by a close rational ratio since in this case the spectrum (\ref{LLq}) is periodic in energy, with periodicity $T_E=C_ebp_0$

In order to estimate the total free energy in the general case, it
is necessary to solve numerically Eq. \ref{LLq}. 
We assume that $C_h=C_ep_0/q_0$ where $p_0$ and $q_0$ are coprime integers \cite{Fo98,Fo05}. Indeed, it is useful, also for computing time reasons, to approximate any real value of $C_h/C_e$ by a close rational ratio since in this case the spectrum (\ref{LLq}) is periodic in energy, with periodicity $T_E=C_ebp_0$
Within each interval $T_E$, there are exactly $p_0+q_0$ discrete solutions
(there is indeed conservation of the number of LLs when $q$
decreases from unity). By successive energy translations, it is
then easy to reproduce all the spectrum. The GS energy $\Delta
E_0$ given by Eq. (\ref{GS}) is however no more well defined, due
to the mixing between electron and hole levels. Indeed, the GS
energy should correspond to the sum of all the quasi-hole energies
below the Fermi energy $E_F(b)$. These energies being not bounded
by below, the GS energy is formally infinite. We propose to
account numerically for this problem by introducing a cutoff
function in the density of states with the necessary condition
that the magnetization should not depend on variations of the
cutoff parameters.

Given a set of LL energies $E_{eh}(q,n)$, $n=0,\cdots,p_0+q_0-1 $,
solutions of (\ref{LLq}) in any energy interval of width $T_E$, we
can introduce a cutoff function $\varphi_c(E)$ such as
$\varphi_c(E)=1$ for $E$ larger than a characteristic energy $E_c$
and equal to $\exp[-c(E-E_c)^{2\delta}]$ for $E\le E_c$, where
$\delta$ is any positive integer greater than 1 (we take
$\delta=4$ in the simulations which gives a very smooth cutoff
function). This function has the required property of conserving
the important physical properties near the Fermi surface and
making in particular the GS energy finite.
 The LL density of states $\rho_c(E)$ takes the following form

\bb\nn
\rho_c(E)&=&b\sum_{n=0}^{p_0+q_0-1}\sum_{k=-\infty}^{\infty}\varphi_c(E)
\delta(E-E_{eh}(q,n)-kT_E)
\ee

Given $E_c$, the parameter $c$ is found to be solution of the equation of conservation
at zero temperature:

\bb\nn
N_{eh}&=&\int_{-\infty}^{E_F(b)} dE \rho_c(E)
\\ \nn
&=&
b\int_{-\infty}^{E_F(b)} dE \sum_{n=0}^{p_0+q_0-1}\sum_{k=-\infty}^{\infty}\varphi_c(E)
\\ \label{cons}
& &\delta(E-E_{eh}(q,n)-kT_E)
\ee

where $N_{eh}$ is, as mentioned before, the total number of quasiparticles in the Canonical Ensemble. Once the parameter $c$ is obtained, we can define for example the GS energy $\Delta E_0$

\bb\nn
\Delta E_0&=&b\int_{-\infty}^{E_F(b)} dE \sum_{n=0}^{p_0+q_0-1}\sum_{k=-\infty}^{\infty}\varphi_c(E)
\\ \label{GSE}
& &E\delta(E-E_{eh}(q,n)-kT_E)
\ee

or the free energy $\Delta F$

\bb\nn
\Delta F&=&-tb\int_{-\infty}^{\infty} dE \sum_{n=0}^{p_0+q_0-1}\sum_{k=-\infty}^{\infty}\varphi_c(E)
\\ \label{freeE} \nn
& &\log[1+\exp\beta(\mu-E)]\delta(E-E_{eh}(q,n)-kT_E)
\\
&+&N_{eh}\mu
\ee

\begin{figure}


\centering \resizebox{\columnwidth}{!}
{\includegraphics*{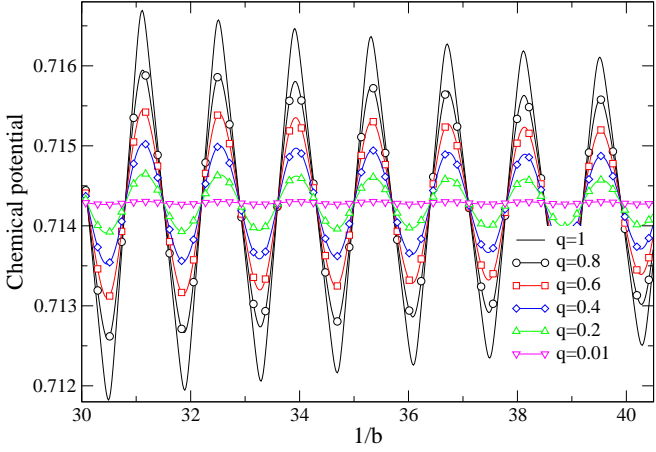}} \caption{\label{muVSq} (color
online) Oscillations of the chemical potential  at $t = 10^{-3}$ for
different values of $q$; $\Delta=1$, m$_e^*$ = 1 and m$_h^*$ =
5/2. }

\end{figure}

The potential $\mu(t,b)$ is calculated from Eq. (\ref{freeE}) by
extremizing the free energy $\partial \Delta F/\partial \mu=0$,
and compelling the magnetization $M_{osc}=-\partial \Delta
F/\partial b$ to be independent of the parameter $E_c$ for $E_c$
far away from the chemical potential or at energies large compared
to the Landau gap. We have checked, for different values of $E_c$
in units of $\Delta$, for example, $E_c=-1,-2,-4$, and for a large
range of fields, that the resulting magnetization does not change.
We choose $N_{eh}$, which is arbitrary, as a multiple of the
characteristic zero field energy density
$m_e^*+m_h^*=(C_e+C_h)/C_eC_h$ times the zero field Fermi energy
$E_F$. For $E_c=-2$, we take in the following numerical
simulations $N_{eh}=8E_F(C_e+C_h)/C_eC_h$. Then, for each value of
the field $b$, the parameter $c$ defined from Eq. (\ref{cons}) is
unique.
 Also, in the case $q=1$, the numerical solution for the magnetization is fully
consistent with
the results of the first section in absence of magnetic breakdown.
In Fig. \ref{muVSq}, the oscillations of the chemical potential $\mu(t,b)$ obviously decrease with $q$ until the total tunneling occurs where only orbits with frequency zero are allowed.

\subsection{Analytical amplitudes for small field}

In this section, we extract the analytical expression for the first harmonics amplitude (corresponding to the frequency $F_0$) in the small $b/t$ regime, and compare it to
the numerical results of the previous section. We first assume that the oscillations can be described by an effective free energy which is constructed by adding reflection amplitudes $q^n$ to the temperature reduction factors in the expression of the Grand Potentials Eqs. \ref{Omega_e} and \ref{Omega_h}, for orbits that circulate $n$ times around Fermi surfaces $S_e$ or $S_h$. In this approximation, it is indeed clear that the quasi-particles orbit
the same surface if the field is small enough, and do not tunnel through the junction.
The effective free energy can then be derived directly from Eqs. (\ref{Omega_e}) and (\ref{Omega_h}) as 

\bb\label{Feff}
\Delta F_{eff}&=&\Omega_e-\Omega_h\simeq
-\frac{\mu^2}{2C_e}-\frac{(\Delta-\mu)^2}{2C_h}
\\ \nn
&+&
\frac{b^2C_e}{2}
\sum_{n=1}^{\infty}\frac{(-q)^n}{\pi^2n^2}R(nm_{e}^*)\cos(2\pi
n\frac{\mu}{C_eb})
\\ \nn
&+&
\frac{b^2C_h}{2}
\sum_{n=1}^{\infty}\frac{(-q)^n}{\pi^2n^2}R(nm_{h}^*)\cos(2\pi
n\frac{\Delta-\mu}{C_hb})
\ee

and the chemical potential is derived as in Eq. \ref{muexact}

\bb \nn \mu&=&E_F+\frac{b}{m^*_e+m^*_h}
\sum_{n=1}^{\infty}\frac{(-q)^n}{\pi n} [ R(nm_{h}^*)\sin(2\pi
n\frac{\Delta-\mu}{C_hb})
\\ \label{mu_q}
&-&R(nm_{e}^*)\sin(2\pi n\frac{\mu}{C_eb}) ] \ee

\begin{figure}
\centering
\resizebox{0.9\columnwidth}{!}
{\includegraphics*{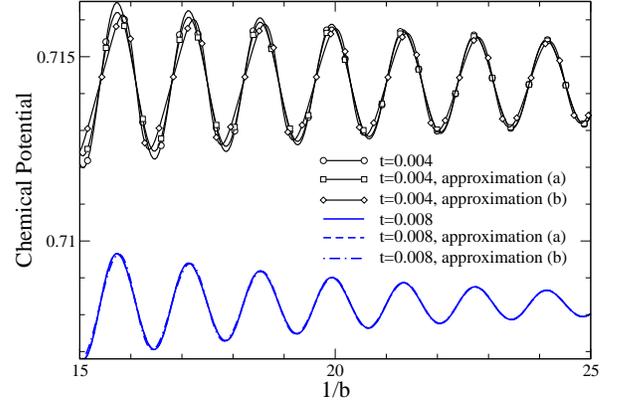}} \caption{\label{AppOsc} (color online)
Comparison between the numerical solution for $\mu$ (Eq. \ref{freeE}) with approximation (a)
from Eq. (\ref{mu_q}), and (b) from Eq. (\ref{mu_q}) with $\mu$ in the sine functions replaced 
by $E_F$. $m_e^*=1$ and $m_h^*=5/2$, and $q=0.6$.
The 2 sets $t=0.004$ and $t=0.008$ are shifted along the vertical axis for clarity.}
\end{figure}

Numerically, we can measure the deviations
between this approximation and the numerical result $\mu=\partial \Delta F/\partial b$, where $\Delta F$ is defined by Eq. (\ref{freeE}), for $m_e^*=1$ and $m_h^*=5/2$ (see Fig. \ref{AppOsc}), and find that Eq. (\ref{mu_q})
is valid for $b/t$ smaller than approximatively $12$ (approximation (a)), and 8 in the case where $\mu$ is replaced by $E_F$ in the right hand side of Eq. (\ref{mu_q}) (approximation (b)), as in Eq. (\ref{mu}) for $q=1$.

The complete derivation of the first harmonic amplitude is done in Appendix A. If we keep the dominant terms in Eq. (\ref{A1}), which are those with the dominant
reduction factor $R(m_{e(h)}^*)$, and discard the other exponentially
small ones $R(nm_{e(h)}^*)$ for $n\ge 2$ or products $R(m_{e(h)}^*)^2$, we obtain

\bb
\amp&\simeq&\frac{2F_0}{\pi}q
\left \{
\frac{R(m_{e}^*)}{m_{e}^*}\frac{J_1(\alpha_e)}{\alpha_e}
+
\frac{R(m_{h}^*)}{m_{h}^*}\frac{J_1(\alpha_h)}{\alpha_h}
\right \}
\ee

where $\alpha_{e(h)}=2m_{e(h)}^*q(R(m_{e}^*)-R(m_{h}^*))/(m_e^*+m_h^*)$ as defined
in Appendix A, and $J_1$ is the Bessel function of order 1. 
In the limit where $\alpha_{e(h)}$ are small, the functions $2J_1(\alpha_{e(h)})/\alpha_{e(h)}$ are very close to unity and we recover the LK
formula:

\bb\label{A1LK}
\amp^{LK}&\simeq&\frac{F_0}{\pi}q
\left \{
\frac{R(m_{e}^*)}{m_{e}^*}
+
\frac{R(m_{h}^*)}{m_{h}^*}
\right \}
\ee

For example, for $q=0.6$, $m_e^*=1$ and $m_h^*=5/2$, $\alpha_e$
vanishes when $b/t$ is either large or small, $\alpha_e$ being always
less than 0.2 elsewhere. At this extremum value, the function $2J(x)/x$ is
approximatively equal to 0.995. Therefore we do not expect, as in
the case $q=1$, a significant deviation from the LK formula at
small $b/t$, as it can be seen in Fig. \ref{A1vsq} in this range
of fields for different values of $q$ strongly different from
unity.

\subsection{LK formula for large field}

In the large $b/t$ limit, the previous approximation is no more
valid (the reduction factors are close to unity). We show
hereafter that there are an infinite number of orbits contributing
to the frequency $F_0$. These orbits, which are described in details in
Appendix B, have large effective masses $m_e(n)=(n+1)m_e^*+nm_h^*$ or
$m_h(n)=nm_e^*+(n+1)m_h^*$, where $n$ is a positive integer. Their
contribution to the total amplitude are therefore negligible at
low field (or high temperature) since their reduction factors are
exponentially small. An example of orbit of mass $m_e(1)$
contributing to frequency $F_0$ is shown in figure \ref{graph2}.

We can however compute the exact amplitude $\amp$ within the LK
theory, and show that there is no discernible difference with the
numerical solution of Eq. (\ref{freeE}) in the Canonical Ensemble.

At low fields, the first harmonics amplitude $\amp$ contains the contribution of
the 2 single orbits of mass $m_e^*$ and $m_h^*$, as in Eq. (\ref{A1LK}).
At higher fields, we can replace the previous quantity by an exact expression

\bb\label{amp1}
\amp^{LK}&=&\frac{F_0}{\pi}
\left (
T_e+T_h
\right )
\ee

where the amplitudes $T_{e(h)}$ includes all the orbits that contribute to the
first harmonics

\bb \nn
T_{e(h)}&=&\frac{q}{m_{e(h)}^*}R(m_{e(h)}^*)
\\ \label{amp2}
&+&\sum_{n=1}^{\infty}
\frac{t_{e(h)}(n)}{m_{e(h)}(n)}R({m_{e(h)}(n)})
\ee

\begin{figure}


\centering \resizebox{\columnwidth}{!}
{\includegraphics*{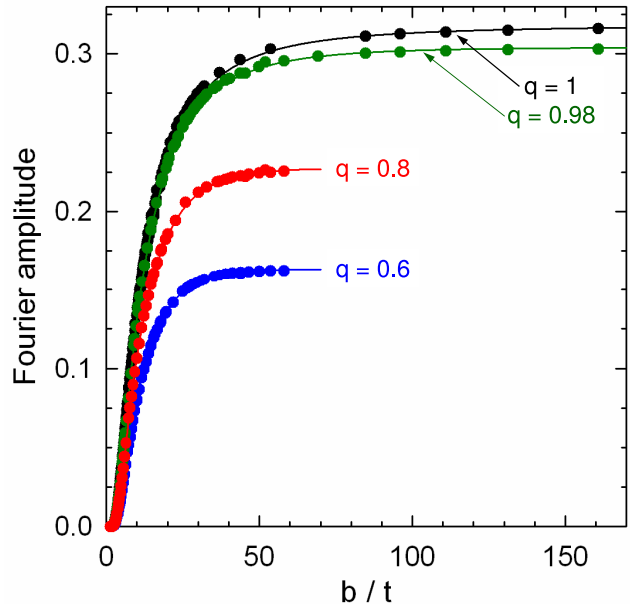}} \caption{\label{A1vsq} (color
online) First harmonics amplitude $\amp$ for different values of
$q$; $\Delta=1$, $m_e^*=1$ and $m_h^*=5/2$. The filled circle
symbols represent the numerical analysis from the free energy
expression (\ref{freeE}), and the lines represent the LK formula
(\ref{amp1}, \ref{amp2}), with elements $t_{e(h)}(n)$ of Eq.(\ref{cnk}) computed up
to $n=10$ from Eq. (\ref{amp3}).}

\end{figure}

\begin{figure}


\centering
\resizebox{0.6\columnwidth}{!}
{\includegraphics*{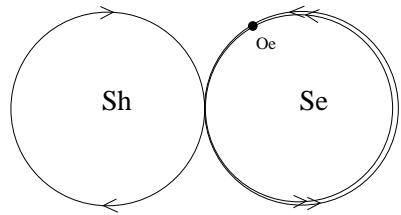}} \caption{\label{graph2} Example of orbit
contributing to first harmonics $F_0$. The effective mass here is $m_e(1)=2m_e^*+m_h^*$.
This orbit can be represented by the operator $\hat Q_e\hat P\hat P $ or
$\hat P\hat P \hat Q_e$ (see Appendix B for details). The point $O_e$ on
surface $S_e$ is crossed twice by the trajectory.}
\end{figure}

The combinatorial coefficients $t_{e(h)}(n)$ defining the total
amplitude for nonequivalent orbits of mass $m_{e(h)}(n)$ are
computed in Appendix B. By definition, we set $t_{e(h)}(0)=q$. For
comparison, we have solved numerically the magnetization from Eq.
(\ref{freeE}) for different values of $q$ and extracted the first
harmonics $\amp$ from Fourier analysis. In Fig. \ref{A1vsq} the
amplitude of the first harmonics $\amp$ is plotted versus $b/t$
for $q=0.6, 0.8, 0.98$ and $1$ (symbols). Since $q$ depends on
$b$ through the Chambers formula, these plots should merely be regarded as the temperature
dependence of $\amp$ at a given magnetic field value. The data are
compared to the predictions of the LK theory (solid lines) given
by Eqs. (\ref{amp1}), (\ref{amp2}) and (\ref{amp3}), for $m_e^*=1$
and $m_h^*=5/2$. The deviations with the LK formula are negligible
in all the $b/t$ range explored, whatever the $q$ value. In
particular, the contributions of the higher mass orbits which
become important in the large field range are still well described
by the LK formula.

\section{Summary and conclusion}

The field-dependent chemical potential oscillations in FS composed of
two compensated electron and hole orbits is strongly damped when
compared to the case of a FS with only electronic orbits
\cite{Al96,Ch02,Ki02,Fo05}. It is even suppressed in the case
where the effective mass of electron (m$^*_e$) and hole (m$^*_h$) band
are the same (assuming the relaxation times $t^*_{e(h)}$ are either
identical or negligibly small). In addition, the LK formula
accounts for the field and temperature dependence of the first
harmonic's dHvA oscillations amplitude in all cases. As for the
amplitude of the second harmonics, it is accounted for by the LK
formula, provided m$^*_e$ and m$^*_h$ are not strongly different.
Besides, as MB develops, the chemical potential oscillations are
further damped and the previous conclusion on the LK validity
still holds for the amplitude of the first harmonic. In this case,
as well as in the general case of compensated electron-hole
orbits, the contributions of an infinite number of orbits to the
main frequency become relevant at high fields, due to the
existence of closed trajectories in the FBZ which have
zero frequency and make the expression of the LK amplitude
rather complex. Within the FS topology that we have considered here, we
have demonstrated that it can nevertheless be computed exactly. We
expect these additional "zero frequency" orbits to be reminiscent 
in other systems as well.

Finally, it should be noticed that the Landau energy spectrum of
1D and 2D periodic networks of electron-hole compensated orbits
(see Figs. \ref{fig_Networks}(b) and (a), respectively), is likely
a set of Landau bands rather than discrete Landau levels as
observed in Figs. \ref{Levelq06me1mh52} and  \ref{Levelq06me1mh1}.
This feature could induce a MB-induced modulation of the density
of states, as observed in non-compensated networks, \cite{Sa97,Fo98, Gv02}
and hence frequency combinations in dHvA oscillatory
spectra. In that respect, the reported method for solving Eq.
\ref{LLq} can be used for the study of such orbits networks.

\appendix

\section{}

From Eq. (\ref{mu_q}), we can define the periodic function

\bb\label{Gfunction}
G(x)=\sum_{n=1}^{\infty}\frac{(-q)^n}{\pi n}
[R(nm_{h}^*)-R(nm_{e}^*)]\sin(2\pi nF_0x)
\ee

where $x=1/b$. The chemical potential Eq. (\ref{mu_q}) becomes $\mu=E_F+bG(x)/(m^*_e+m^*_h)$.
Replacing $\mu$ in (\ref{Feff}) by the previous expression, we compute the oscillating part
of the magnetization $M_{osc}=-\partial \Delta F_{eff}/\partial b=x^2\partial \Delta F_{eff}/\partial x$:

\bb\label{Meff-1}
M_{osc}&\approx& -\frac{C_eC_h}{(C_e+C_h)}G(x)G'(x)
\\ \nn
&+&\frac{1}{2}
\sum_{n=1}^{\infty}\frac{(-q)^n}{\pi^2n^2}
\left [
C_eR(nm_{e}^*)\frac{\partial}{\partial x} \Re \exp(2i\pi n\frac{\mu}{C_e}x)
\right .
\\ \nn
&+& \left .
C_hR(nm_{h}^*)\frac{\partial}{\partial x} \Re \exp(2i\pi n\frac{\mu-\Delta}{C_h}x)
\right ]
\ee

We also define the Fourier coefficients $B_{e(h)}(n,m)$ as

\bb\label{FourierB}
e^{2i\pi n w_{e(h)}G(x)}=\sum_{m=-\infty}^{+\infty}B_{e(h)}(n,m)e^{2i\pi mF_0x}
\ee

where $w_{e(h)}=C_{h(e)}/(C_e+C_h)$. Then, we perform the derivatives in
expression (\ref{Meff-1}) by noticing for example that

\bb\nn
& &\frac{\partial}{\partial x} \Re \exp(2i\pi n\frac{\mu}{C_e}x)
=\Re \left [ \sum_{m=-\infty}^{+\infty}B_{e}(n,m)\right .
\\ \label{der-1}
& &\left .
\times 2i\pi(m+n)F_0\exp(2i\pi(m+n)F_0x)
\right ]
\ee

and

\bb\nn
& &\frac{\partial}{\partial x} \Re \exp(2i\pi n\frac{\mu-\Delta}{C_h}x)
=\Re \left [ \sum_{m=-\infty}^{+\infty}B_{h}(n,m)\right .
\\ \label{der-2}
& &\left .
\times 2i\pi(m-n)F_0\exp(2i\pi(m-n)F_0x)
\right ]
\ee

For extracting the first amplitude, we select the integers such as $m+n=\pm 1$
in (\ref{der-1}) and $m-n=\pm 1$ in (\ref{der-2}). In the product $G(x)G'(x)$ of 
expression (\ref{Meff-1}),
the terms can be rearranged noticing that $\cos(2\pi mF_0x)\sin(2\pi nF_0 x)=
[\sin(2\pi(n+m)F_0 x)+\sin(2\pi(n-m)F_0 x)]/2$ with $m,n\ge 1$. For the
first harmonic, only terms $n-m=\pm 1$ contribute.
 A further approximation is to keep the first term of the series (\ref{Gfunction}),
i.e. $G(x)\approx -q(R(m_{h}^*)-R(m_{e}^*))\sin(2\pi F_0x)/\pi$, in (\ref{FourierB}),
so that the coefficients $B_{e(h)}(n,m)$ can be computed exactly, from the explicit relation

\bb\nn
& &\exp\left [
2inw_{e(h)}q(R(m_{e}^*)-R(m_{h}^*))\sin(2\pi F_0x)
\right ]
=
\\
& &\sum_{m=-\infty}^{\infty}J_{m}(n\alpha_{e(h)})\exp(2i\pi mF_0x)
\ee

where $J_m$ is the Bessel function of order $m$ and
$\alpha_{e(h)}=2w_{e(h)}q(R(m_{e}^*)-R(m_{h}^*))$. In this
case it is easy to identify $B_{e(h)}(n,m)=J_m(n\alpha_{e(h)})$.
After some algebra, we find that the amplitude $\amp$ for the
first harmonic $F_0$ is given by the expression, in the large $t/b$ limit

\bb\nn
\frac{\pi\amp}{F_0}&\approx&\sum_{n=1}^{\infty}2\frac{(-q)^n}{n}
\left \{
(-1)^nC_eR(nm_{e}^*)\frac{J_n(n\alpha_e)}{n\alpha_e}
\right .
\\ \nn
&-&
\left .
C_hR(nm_{h}^*)\frac{J_n(n\alpha_h)}{n\alpha_h}
\right \}+\frac{C_eC_h}{C_e+C_h}\times
\\ \label{A1}
& &\sum_{n=1}^{\infty}\frac{1}{n}
[R(nm_{h}^*)-R(nm_{e}^*)]
\times
\\ \nn
& &\left \{
[R((n-1)m_{h}^*)-R((n-1)m_{e}^*)]q^{2n-1}
\right .
\\ \nn
& &\left .
-[R((n+1)m_{h}^*)-R((n+1)m_{e}^*)]q^{2n+1}
\right \}
\ee

\section{}

In this appendix, we compute the amplitude of the first
hamonics in the LK theory for any given reflection amplitude
$q$. In the lowest approximation, the contribution to the amplitude
is given by one single orbit around the electron or hole band (see Fig. \ref{FSMB}).
Each contributes with a mass equal to $m_e^*=1/C_e$ and
$m_h^*=1/C_h$ respectively and a reflection amplitude $q$ at the junction
point. At finite temperature, we add a reduction factor $R(m_{e}^*)$ or $R(m_{h}^*)$
so that we obtain (with a $F_0$ factor overall) Eq. (\ref{A1LK}).

However, there are other contributions to the $F_0$ harmonics,
coming from more complex orbits. For example in Fig. \ref{graph2}, from a starting
point $O_e$ near the junction on the Fermi surface $S_e$, we can
reach the junction point and go through to the other band $S_h$,
then complete an entire orbit and come back through the same
junction to the band $S_e$, and finally perform 2 orbits around
$S_e$ until we reach $O_e$ again. The
frequency will be proportional to $2S_e-S_h$. For a compensated
system, this frequency is just $F_0$ since $S_e=S_h$. The total
mass is however proportional to the derivative of $2S_e-S_h$ with
respect to the energy, which is $2m_e^*+m_h^*$ (the derivative of
$S_h=2\pi m_h^*(\Delta-E)$ is indeed negative). Therefore we
should take into account all possible trajectories. This can be done
exactly by summing up all the contributions for a given mass
$m_e(n)=(n+1)m_e^*+nm_h^*$ ($n\ge 0$).
 As before we start from a point $O_e$ ($O_h$) on the surface $S_e$ ($S_h$) and turn until we reach
the junction. Here we note $\hat Q_e$ ($\hat Q_h$) the reflection operator which keep
the quasiparticle on the surface $S_e$ with amplitude $q$ ($S_h$ respectively),
and $\hat P$ the operator which take the quasiparticle through the junction (with amplitude $ip$). Their electron-hole representations are given by the matrices:

\bb
\hat Q_e=\hat Q_h=
\left (
\begin{array}{cc}
q & 0
\\
0 & q
\end{array}
\right )
,\;\;
\hat P=
\left (
\begin{array}{cc}
0 & ip
\\
ip & 0
\end{array}
\right )
\ee

In the simplest case, the orbit is described only by the operator or graph $\hat Q_e$,
whereas in the second example, the orbit is described by the graph $\hat P\hat P \hat Q_e$.
The orbit $\hat Q_e\hat P\hat P $ is equivalent and has the same mass $2m_e^*+m_h^*$.
Indeed, the orbit goes trough the same point $O_e$ twice, and therefore there are 2
different ways of writing the graph above, depending on which branch of the orbit
we place the point $O_e$. Equivalent graphs are described by a cyclic permutation
of their operator elements.
We also observe that multiple orbits $(\hat P\hat P)^n$ have zero frequency and mass $n(m_e^*+m_h^*)$.
We will note in the following $t_e(n)$ ($t_h(n)$) the sums of the amplitudes of all the trajectories with the same mass $m_e(n)$ ($m_h(n)=nm_e^*+(n+1)m_h^*$ respectively), which are constructed starting from an arbitrary point $O_e$ (or $O_h$ respectively). $T_e$ ($T_h$) will
be the total amplitude for the frequency $F_0$ ($-F_0$) starting from point $O_e$ ($O_h$) on the surface $S_e$ ($S_h$), see Eqs. (\ref{amp1},\ref{amp2})


$T_h$ is computed in the same way as $T_e$ by replacing the label $e$ in all the
quantities by $h$. For equal masses $m_e^*=m_h^*$, we have clearly $T_e=T_h$.
 For example, the orbits and amplitudes contributing to the mass $m_e(1)=2m_e^*+m_h^*$ are

\bb\nn
\hat P\hat P\hat Q_e\sim \hat Q_e\hat P\hat P
\rightarrow (-p^2)q,
\ee

This gives $t_e(1)=(-p^2)q$. The 2 operators above correspond to the same graph
by a cyclic permutation of their elements $\hat P\hat P$ and $\hat Q_e$ and the
sign $\sim$ is understood as an equivalence between identical graphs.
In the next case, for the mass $m_e(2)=3m_e^*+2m_h^*$ we have the possibilities

\bb\nn
\hat P\hat Q_h\hat P\hat Q_e^2\sim \hat Q_e\hat P\hat Q_h\hat P\hat Q_e\sim
\hat Q_e^2\hat P\hat Q_h\hat P
\rightarrow (-p^2)q^3,
\\ \nn
\hat P\hat P\hat P\hat P\hat Q_e\sim \hat P\hat P\hat Q_e\hat P\hat P\sim \hat Q_e\hat P\hat P\hat P\hat P
\rightarrow (-p^2)^2q
\ee

we obtain $t_e(2)=(-p^2q^3+p^4q)$. In the general case, we can construct all the possible
orbits that contribute to the mass $m_e(n)$ by representing them as multiple
products of elementary operators

\bb\label{graph}
\hat Q_e^{n_1}\hat P\hat Q_h^{m_1}\hat P\hat Q_e^{n_2}\hat P\hat Q_h^{m_2}\hat P..
\hat Q_e^{n_k}\hat P\hat Q_h^{m_k}\hat P
\ee

where $n_i\ge 0$ and $m_i\ge 0$ are positive integers. The constraints imposed
by the mass on theses integers and $k$ are

\bb\label{constraint}
\sum_{i=1}^{k}n_i=n-k+1,\;\sum_{i=1}^{k}m_i=n-k,\;1\le k\le n
\ee

There are also $k$ cyclic permutations of the graph (\ref{graph}) above, corresponding
to moving elements $\hat Q_e^{n_i}\hat P\hat Q_h^{m_i}\hat P$ to the right or to the left.
For example, an equivalent graph would be

\bb
\hat Q_e^{n_2}\hat P\hat Q_h^{m_2}\hat P..\hat Q_e^{n_k}
\hat P\hat Q_h^{m_k}\hat P\hat Q_e^{n_1}\hat P\hat Q_h^{m_1}\hat P
\ee

The total amplitude $t_e(n)$ for a given $n$ can be written as

\bb\label{amp3}
t_e(n)&=&\sum_{k=1}^nc_{n,k}(-p^2)^k
q^{2(n-k)+1}
\ee

where the coefficients $c_{n,k}$ enumerate the number of all
nonequivalent graphs having the same mass, up to cyclic
permutations. It corresponds to all possible sets of $n_i$ and
$m_i$ representing graphs (\ref{graph}) with the constraint
(\ref{constraint}), divided by the number of cyclic
permutations $k$. The total amplitude $t_e(n)$ can be written more
precisely as

\bb\nn
t_e(n)&=&\sum_{k=1}^n\sum_{n_i,m_i\ge 0}
q^{\sum_{i=1}^{k}n_i+\sum_{i=1}^{k}m_i}
(-p^2)^k\times
\\
& &\frac{1}{k}\delta_{\sum_{i=1}^{k}n_i,n-k+1}
\delta_{\sum_{i=1}^{k}m_i,n-k}
\ee

The Kronecker functions $\delta_{k,0}$ can be represented by integrals
$\delta_{k,0}=\int_0^1\exp(2i\pi kx)dx=\oint dz/(2i\pi z)\times z^{k}$, and
the last expression becomes

\bb
& & t_e(n)=\sum_{k=1}^n\frac{(-p^2)^k}{k}\oint \frac{dz}{2i\pi z}\oint \frac{dz'}{2i\pi z'}
\\ \nn
& &\left (
\prod_{i=1}^{k}\sum_{n_i=0}^{\infty}q^{n_i}z^{n_i}
\right )z^{k-n-1}
\left (
\prod_{i=1}^{k}\sum_{m_i=0}^{\infty}q^{m_i}z'^{m_i}
\right )
z'^{k-n}
\ee

Summing up the series, we obtain

\bb\nn
t_e(n)&=&\sum_{k=1}^n\frac{(-p^2)^k}{k}\oint \frac{dz}{2i\pi}
\frac{z^{k-n-2}}{(1-qz)^{k}}
\oint \frac{dz'}{2i\pi}
\frac{z'^{k-n-1}}{(1-qz')^{k}}
\ee

Applying the residue theorem, it is easy to show that

\bb
\oint \frac{dz}{2i\pi}
\frac{z^{k-n-2}}{(1-qz)^{k}}
=\left (^{n}_{k-1}\right )q^{n-k+1}
\ee

and

\bb
\oint \frac{dz'}{2i\pi}
\frac{z'^{k-n-1}}{(1-qz')^{k}}
=\left (^{n-1}_{k-1}\right )q^{n-k}
\ee

where $\left (^{n}_{k}\right )=n!/k!(n-k)!$ are the binomial coefficients. We obtain finally

\bb\label{cnk}
c_{n,k}=\frac{1}{k}\left (^{n}_{k-1}\right ) \left (^{n-1}_{k-1}\right )
\ee

The values of the coefficients $c_{n,k}$ up to $n=7$ are given in Table 1.

\begin{table}[ht]
\caption{Values of the coefficients $c_{n,k}$ which represent
the number of non-equivalent orbits for a given mass $(n+1)m_{e(h)}^*+nm_{h(e)}^*$
with $2k$ breakdowns, $1\le k\le n$. }
\centering
\begin{tabular}{| c | c c c c c c c | c |}
\hline
$_n\backslash^k$ & 1 & 2 &  3 & 4 & 5 & 6 & 7 & Mass
\\
\hline
1 & 1 & & & & & & & $2m_{e(h)}^*+m_{h(e)}^*$
\\
2 & 1 & 1 &  & & & & & $3m_{e(h)}^*+2m_{h(e)}^*$
\\
3 & 1 & 3 & 1 & & &  & & $4m_{e(h)}^*+3m_{h(e)}^*$
\\
4 & 1 & 6 & 6 & 1 &  & & & $5m_{e(h)}^*+4m_{h(e)}^*$
\\
5 & 1 & 10 & 20 & 10 & 1  & & & $6m_{e(h)}^*+5m_{h(e)}^*$
\\
6 & 1 & 15 & 50 & 50 & 15 & 1  &  & $7m_{e(h)}^*+6m_{h(e)}^*$
\\
7 & 1 & 21 & 105 & 175 & 105 & 21 & 1 & $8m_{e(h)}^*+7m_{h(e)}^*$
\\
\hline
\end{tabular}
\end{table}

We also observe by symmetry that $t_h(n)$ is equal to $t_e(n)$
since this geometric coefficient does not depend explicitely on masses.

\bibliographystyle{unsrt}
\bibliographystyle{apsrev}
\bibliography{biblioCOMP-MET1}

\begin{thebibliography}{26}
\bibitem[{fn1()}]{fn1}
\bibinfo{note}{The two electronic orbits considered in Fig. 1 have different
  frequencies, namely f$_0$ / f$_1$ = 2 / 3. In the case where f$_0$ = f$_1$,
  the chemical potential oscillations are even larger than reported in Fig. 1
  since the field-dependent contributions of the two orbits are in phase.}
\end{thebibliography}

\end{document}